\documentclass{article}
\usepackage{hiph-preprint}
\usepackage{amsmath}
\usepackage{epsfig}
\volnumber{22} \issuenumber{1} \edyear{2005}                             %|
\frompage{000} \topage{000}                                              %|
\recrevdate{28 Ocotober 2005}                                              %|
\def\rightmark{Jet-medium interactions: wakes in the QCD medium}
\def\leftmark{J. Ruppert}
 
\title{Jet-medium interactions: wakes in the QCD medium} 
\authors{ 
{J\"org Ruppert$^{1}$  %
}\\[2.812mm]
{\normalsize
\hspace*{-8pt}$^1$ Department of Physics, Duke University, P.O. Box 90305, Durham, 27707, NC, USA}}
\abstract{We calculate the wake induced in a hot, dense QCD medium by a fast parton in the framework of linear response theory. We discuss two different scenarios: (i) a weakly coupled quark gluon plasma (pQGP) described by hard-thermal loop (HTL) perturbation theory and (ii) a scenario where the plasma has the properties of a quantum liquid. We show that a Mach cone can appear in the second scenario, but not in the first one.}

 
\makeindex
\begin{document}
 
\maketitle

\section{Introduction}\label{intro}

Jet quenching of QCD jets created in relativistic heavy ion collisions has been identified as an important indicator for the creation of a quark-gluon plasma \cite{Gyulassy:1990ye}. It is extensively studied theoretically and experimentally at RHIC. 

The main emphasis in theoretical studies has been on the description of the 
radiative energy loss which the leading parton suffers when traversing 
the medium. Several of this year's quark matter contributions addressed also these questions \cite{Salgado:2005pr}.

The main emphasis of the work I want to focus on here is another aspect
of the in-medium jet physics, namely the fact that a charged particle with high momentum traveling through a medium induces a wake of current density, charge density and (chromo-)electric and magnetic field configurations. This wake reflects significant properties of the medium's response to the jet. It is shown that under certain circumstances, namely if the QCD medium behaves like a quantum liquid,  this wake can exhibit Mach cone like structures. 

In this contribution I focus on the theoretical aspects of the calculation of this wake which I have done with Berndt M\"uller  \cite{Ruppert:2005uz}. For a more general discussion of jet medium interactions and for the discussion of possibly observable consequences of the outlined calculation I refer the reader to this Quark Matters focus talk on the subject \cite{Ruppert:2005ic} and the study \cite{Renk:2005si}. 

We applied methods of linear response theory to a system of a relativistic color charge traveling through a QCD plasma. The calculation is restricted to only one weak, external current. In this framework quantum and non-ablian effects are included indirectly via the dielectric functions, $\epsilon_L$ and $\epsilon_T$. Linear response theory implies that the deviations from equilibrium are small enough that they will not, in turn, modify the dielectric function. For simplicity the medium is assumed to be homogenous and isotropic, finite size effects and retardation effects \cite{Peigne:2005rk} are disregarded here.

\section{Formalism}

For an isotropic and homogenous medium the dielectric tensor can be decomposed into its longitudinal and transverse function. Using Maxwell's equation and the continuity equation in momentum space the total chromoelectric field $\vec{E}^a_{\rm tot}$ in the QCD plasma is related to the external current $\vec{j_{\rm ext}}$ via \cite{Ichimaru}:
 \begin{eqnarray} \label{EtotTOJ}
 \left[\epsilon_L {\cal P}_{L} + \left(\epsilon_T -\frac{k^2}{\omega^2}\right) {\cal P}_{T}\right] \vec{E}^a_{\rm tot}(\omega,\vec{k})=\frac{4 \pi}{i \omega} \vec{j}^a_{\rm ext} (\omega,\vec{k}) .
 \end{eqnarray}
This has non-trivial solutions only, if the determinant constructed from the elements of the tensor vanishes.
From this dispersion relations for the longitudinal and transverse dieelectric functions can be derived \cite{Ichimaru}: $\epsilon_L=0,\epsilon_T=(k/\omega)^2$.
These equations determine the longitudinal and transverse plasma modes. 

The color charge density induced in the wake by the external charge distribution is:
\begin{eqnarray} \label{charge}
\rho_{\rm ind}=\left(\frac{1}{\epsilon_L}-1\right)\rho_{\rm ext}.
\end{eqnarray}

A direct relation between the external and the induced current can be derived:
\begin{eqnarray} \label{current}
\vec{j}^a_{\rm ind}=\left[\left(\frac{1}{\epsilon_L}-1\right){\cal P}_L + \frac{1-\epsilon_T}{\epsilon_T-\frac{k^2}{\omega^2}} {\cal P}_T\right]\vec{j}^a_{\rm ext}.
\end{eqnarray}
At this point we specify the current and charge densities associated with a color charge as Fourier transform of a point charge moving along a straight-line trajectory with constant velocity $\vec v$:
$\vec{j}^a_{\rm ext}=2\pi q^a \vec{v} \delta(\omega-\vec{v} \cdot \vec{k}),\vec{\rho}^a_{\rm ext}=2\pi q^a \delta(\omega-\vec{v} \cdot \vec{k})$. where $q^a$ is its color charge defined by $q^a q^a = C \alpha_s$ with the strong coupling constant $\alpha_s=g^2/4\pi$ and the quadratic Casimir invariant $C$ (which is either $C_F=4/3$ for quarks or $C_A=3$ for a gluon). In this simplified model one disregards changes of the color charge while the particle is propagating through the medium by fixing the charge's orientation in color space \cite{Weldon,Thoma}.

\section{Results and discussion: charge wakes in a QGP}

We discuss two qualitatively different scenarios, in the first one we assume that the plasma is in the high temperature regime where the gluon self-energy can be described using the leading order of the high temperature expansion, which is equivalent to the HTL approximation. The dielectric functions are therefore gauge invariant and plasma modes can only appear in the time-like sector of the $\omega,k$ plane. 
In this scenario the wake induced by a partonic jet reveals a charge and current density profile in which no Mach cones appear, but the charge carries a screening color cloud along with it. Fig. 1 in the first paper of \cite{Ruppert:2005uz}  shows the charge density of a colored parton traveling with $v=0.99~c$ in cylindrical coordinates. 

In the second scenario we investigate what happens if the plasma is in a regime where it exhibits properties of a quantum liquid. Since there is a lack of first principle calculations of how to calculate the dielectric function in such a regime. We use a simple model. 
Nonetheless, this simplified model is constructed in such a way that it allows for a general conclusion quite independent from the exact form of the dielectric functions. To be specific it is assumed that the longitudinal plasmon of the quantum liquid like plasma has the following Bloch \cite{Bloch} dispersion relation  $\omega=\sqrt{c_s^2 k^2 + \omega_p^2}$, where $\omega_p$ denotes the plasma frequency and $c_s$ the speed of colored sound in units of $c$. It is assumed that a critical momentum $k_c$ exists which separates the regimes of collective and single particle modes in the quantum liquid, therefore this collective mode exist for $k\le k_c$ only.
Such a plasmon dispersion relation is realized by the following longitudinal dielectric function:
\begin{eqnarray}
\epsilon_L=1+\frac{\omega_p^2/2}{c_s^2k^2-\omega^2+\omega_p^2/2}\,\,(k\le k_c)\,
\end{eqnarray}
Note that this differs from the classical, hydodynamical function of Bloch since 
the latter is singular at small $k$ and $\omega$.
The Bloch (like) plasmon mode extends for $k>\omega_p/\sqrt{1-c_s^2}$ into the space-like region of the $\omega,k$ plane. Notice that this is different from the high-temperature plasma, where longitudinal and transverse plasma modes only appear in the time-like region $|\omega/k|>1$. The qualitative properties of the color wake can in the case of a subsonically traveling jet expected to be analogous to those of the high temperature plasma, namely that the charge carries only a screening color cloud with it and Mach cones do not appear. If the jet is traveling supersonically the situation changes and the modes with intermediate phase velocities can be excited. Mach cones can appear. 
To illustrate this we consider a speed of colored sound of $c_s=1/\sqrt{3}$ and calculated the charge density for a colored particle traveling with $v/c=0.99>c_s$, see Fig. 1.

\section{Conclusion}

We calculated the charge density wake induced by a parton travelling through the QCD medium. It is shown that if the plasma can be described in the high temperature approximation (first scenario) the wake exhibits properties of a co-moving screening cloud. In the second scenario we investigate the question what happens if the plasma has the properties of a quantum liquid. It is shown that Mach cones can appear if the parton is traveling supersonically and a longitudinal plasmon extends in the space-like region of the $\omega,k$-plane.

\begin{figure}
 \par\resizebox*{!}{0.20\textheight}{\includegraphics{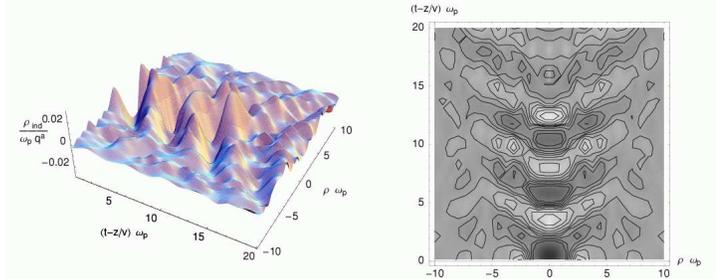}} \par{}
\caption{(a) Spatial distribution of the induced charge density from a  jet with high momentum and fixed color charge $q^a$ that is traveling with 
$v=0.99c>u=\sqrt{1/3}c$. (b) Plot showing equi-charge lines in the density distribution for the situation in (a). 
\label{figure31}}
\end{figure}

\section*{Acknowledgments}
I thank B.~M\"{u}ller as a collaborator and T. Renk for discussions.
This work was supported by DOE grant DE-FG02-96ER40945 and a Feodor Lynen Fellowship of the Alexander von Humboldt Foundation.

\vfill\eject
\end{document}